%% file: main.tex
\documentclass[aps,prl,twocolumn,amsmath,amssymb,floatfix,longbibliography]{revtex4-2}
	
\usepackage{color}
\usepackage{mathrsfs}
\usepackage{graphicx}
\usepackage{dcolumn}
\usepackage{bm}
\usepackage{hyperref}
\usepackage{amsmath,amsfonts,amssymb,ulem}
\usepackage{epstopdf}
\usepackage{xcolor}

\newcommand{\re}[1]{{\color{black}#1}}

\usepackage{graphicx}
\begin{document}

\title{Electron  polarization in ultrarelativistic plasma current filamentation instabilities}
\author{Zheng Gong}
\email[]{gong@mpi-hd.mpg.de}
 \affiliation{Max-Planck-Institut f\"{u}r Kernphysik, Saupfercheckweg 1, 69117 Heidelberg, Germany}
\author{Karen Z. Hatsagortsyan}
\email[]{k.hatsagortsyan@mpi-hd.mpg.de}
 \affiliation{Max-Planck-Institut f\"{u}r Kernphysik, Saupfercheckweg 1, 69117 Heidelberg, Germany}
\author{Christoph H. Keitel}
 \affiliation{Max-Planck-Institut f\"{u}r Kernphysik, Saupfercheckweg 1, 69117 Heidelberg, Germany}

\date{\today}

\begin{abstract}
Plasma current filamentation of an ultrarelativistic electron beam impinging on an overdense plasma is investigated, with emphasis on radiation-induced electron polarization. Particle-in-cell simulations provide the classification and in-depth analysis of three different regimes of the current filaments, namely, the normal filament, abnormal filament, and quenching regimes. We show that electron radiative polarization emerges during the instability along the azimuthal direction in the momentum space, which  significantly varies across the regimes. We put forward an intuitive Hamiltonian model to trace the origin of the electron polarization dynamics. In particular, we discern the role of nonlinear transverse motion of plasma filaments, which induces asymmetry in radiative spin flips, yielding an accumulation of electron polarization. 
Our results break the conventional perception that quasi-symmetric fields are inefficient for generating radiative spin-polarized beams, suggesting the potential of electron polarization as a source of new information on laboratory and astrophysical plasma instabilities.

\end{abstract}

\maketitle
Current filamentation instability (CFI)~\cite{weibel1959spontaneously,Fried1959,lee1973electromagnetic}, triggered by the interpenetration of counterstreaming plasma flows, fragments the driving beam into narrow dense filaments and thereby amplifies self-generated magnetic fields~\cite{califano2001fast,silva2003interpenetrating,bell2004turbulent,califano2006three,bret2008exact}. It is crucial in regulating various plasma phenomena. Microscopically, CFI modifies the electron energy deposition in inertial confinement fusion~\cite{honda2000collective,gremillet2002filamented,sentoku2003anomalous,macchi2003fundamental,li2021nanoscale}, constrains the accelerating gradient of wakefield accelerators~\cite{huntington2011current,allen2012experimental,claveria2022spatiotemporal}, and magnifies magnetic fields in a nonlinear stage following the saturation of the linear Weibel instability~\cite{peterson2021magnetic}.
In the astrophysical world, CFI can catalyze the supernova remnant collisionless shocks~\cite{bell1978acceleration,spitkovsky2008particle,sironi2015relativistic,marcowith2016microphysics,ruyer2016analytical,lemoine2019physics}, instigate stochastic acceleration in turbulent reconnection~\cite{che2011current,hoshino2012stochastic,matsumoto2015stochastic,bohdan2020kinetic}, and reshape the afterglow radiation following gamma-ray bursts~\cite{gamma_ray_burst,medvedev1999generation,milosavljevic2006weibel,ardaneh2015collisionless}.
The latest investigations reveal that CFI facilitates the interpretation of Saturn’s bow shock transition~\cite{Magnetic_field_amp_in_shock_2021} and coherent emission of fast radio bursts~\cite{sironi2021coherent}. The advancement of laboratory astrophysical platforms~\cite{remington2006experimental,lebedev2019exploring,takabe2021recent,meuren2020seminal} will promote further in-depth experimental study of CFI dynamics~\cite{fox2013filamentation,RRD_bulanov2015,Weibel_NP2015_huntington,ruyer2020growth,fiuza2020electron}.

The radiation of ultrarelativistic electrons inside the CFI can lead to generation of compact high-brilliance gamma-rays~\cite{benedetti2018giant} and  copious e$^+$e$^-$ pairs~\cite{nerush2017weibel}.
State-of-the-art techniques of compressed energetic beams such as FACET-II~\cite{yakimenko2019prospect,yakimenko2019facet} will further foster these processes.
Recently, it has been recognized that electrons can be spin-polarized in symmetry-broken magnetic fields~\cite{Seipt_2018,li2019ultrarelativistic,Chen_2019,Seipt_2019,Wan_2020,Guo_2020,gong2021retrieving} due to radiative spin flips, intrinsically accompanying gamma photon emissions~\cite{Sokolov_1968,Baier_1967,Baier_1972}. 
One may ask whether the electrons can be radiatively spin-polarized in such an ultrarelativistic CFI in spite of quasi-symmetric fields and how the features of electron \re{spin polarization (SP)} are correlated with its underlying mechanisms.
Answering these questions will bring to light new details of pertinent plasma instabilities using the SP information.

In this letter, we investigate the dynamics of ultrarelativistic plasma CFI, employing  electron spin resolved particle-in-cell simulations. The classification into three different CFI regimes is introduced based on the distinct collective behavior and radiative spin-flip mechanisms: \re{normal filament (NF), abnormal filament (ANF), and quenching regimes}. We indicate the different topological structures of filaments in the transverse plane in NF and ANF regimes, which results in different filament merging dynamics. The latter has a direct impact on the spontaneous SP of the beam electrons, which  are spin-polarized along the azimuthal direction in momentum space. 
While the electron SP is influenced by the strength of the magnetic fields, the nonlinear transverse motion of the current filaments is found to be vital for the effective accumulation of net SP.
The latter generally appears in the ANF scenario with topologically connected beam electron filaments. In the NF situation, however, the electron SP ratio is weakened by the compensation of the nearly symmetric radiative spin flips. 
The correlation between the emerging SP and the collective behaviors presented here enables decoding CFI-induced scenarios via polarization detection.

\begin{figure}
\includegraphics[width=0.45\textwidth]{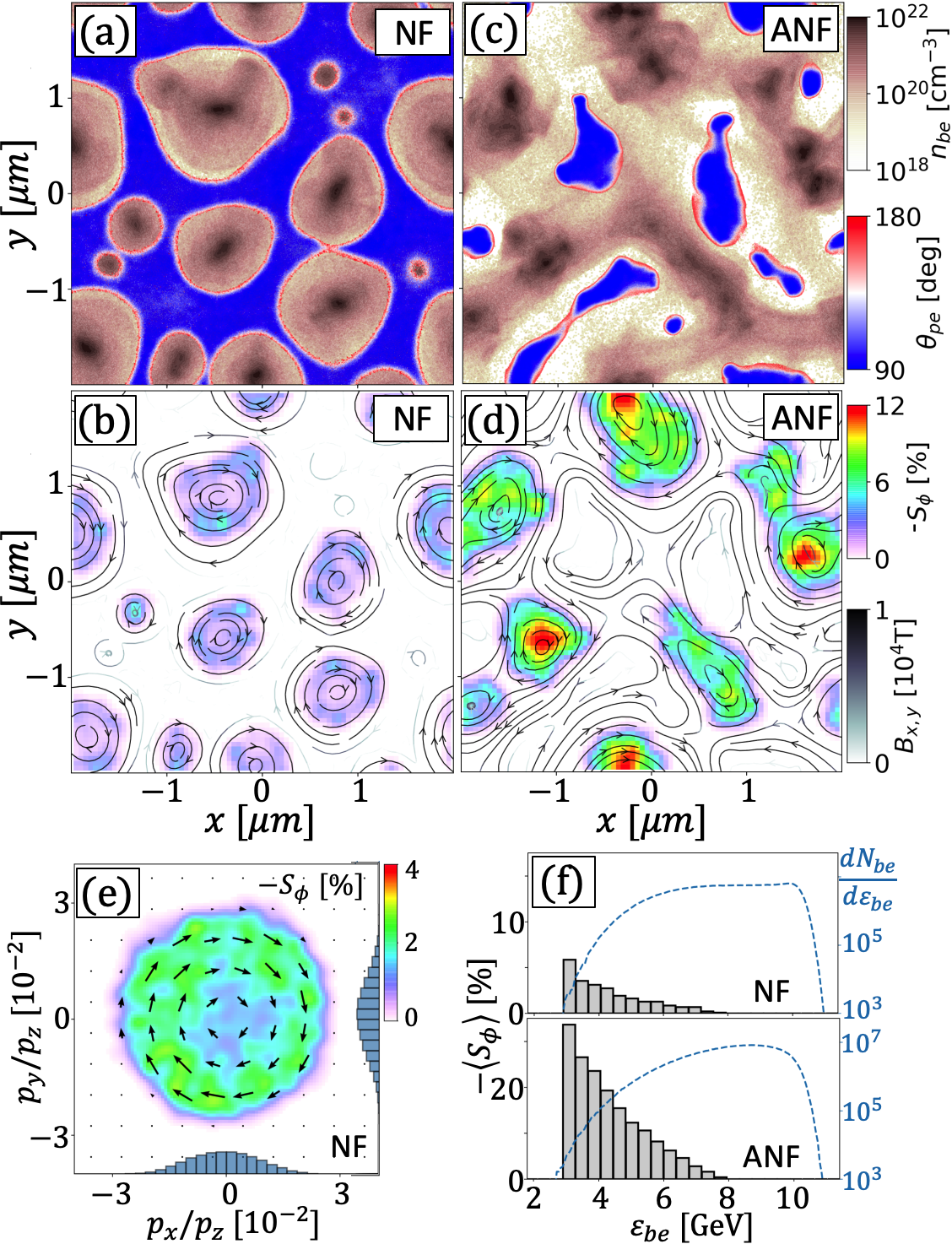}
\caption{(a)(c) Beam electron density $n_{be}$ and the plasma electrons' polar angle $\theta_{pe}=\arctan(p_{\perp},p_{z})$. (b)(d) electron SP ratio $S_\phi$, where the streamlines denote the magnetic field $B_{x,y}$. (a)(b) and (c)(d) correspond to the NF and ANF cases, respectively. (e) SP ratio $S_\phi$ in $(p_y/p_z,p_x/p_z)$ space, where the arrows denote the SP direction and the blue bars show the electron number $dN_{be}/d(p_{x,y}/p_z)$. (f) $\langle S_\phi\rangle$ and $dN_{be}/d\varepsilon_{be}$ vs $\varepsilon_{be}$ for the NF and ANF cases.} 
\label{fig:general}
\end{figure}

In 2D particle-in-cell simulations, the domain in $(x,y)$ space has a dimension of $4\mu m\,\times\,4\mu m$, with a cell size of $\Delta x=\Delta y=1/256\,\mu m$. An ultrarelativistic electron beam with density $n_{be}=5\times10^{20}\,\mathrm{cm}^{-3}$ and energy $\varepsilon_{be}\approx 10 \mathrm{GeV}$ (available at FACET-II in the near future~\cite{yakimenko2019facet}) is initialized to propagate along the $+z$ direction. A background plasma electron flow with a density $n_{pe}$ and velocity $v_z\sim (n_{be}/n_{pe})c$ ($c$ is the speed of light) is set to neutralize the current density at the initial time.
The three main examples with $n_{pe}=10^{22}$, $3\times10^{22}$, and $5\times10^{22}\,\mathrm{cm}^{-3}$ correspond to regimes referred to as ANF, NF, and quenching, respectively.
The ions, with charge $Z_i=1$, mass $m_i=1836m_e$, and density $n_{pi}$, are stationary to neutralize the charge density at the beginning. The computational area is filled with 20 macro-particles for each species per cell. The models of the spin precession governed by the Thomas-Bargmann-Michel-Telegdi equation~\cite{thomas1927kinematics,bargmann1959precession} and the radiative spin flips are implemented in the EPOCH code~\cite{arber2015contemporary,SM} via spin resolved \re{quantum electrodynamical} probabilities \cite{Chen_2022}, using the instantaneous spin quantization axis~\cite{li2020production}.

In the NF regime ($n_{pe}=3\times10^{22}\mathrm{cm}^{-3}$), ultrarelativistic beam electrons are pinched into multiple filaments like a rod array while the background electrons fill up the residual area to encompass the beam filaments [Fig.~\ref{fig:general}(a)], which is similar to the previously measured magnetic tube array structure~\cite{B_tube_array_zhou2018self,jung2005study,quinn2012weibel} so that it is termed as \textit{``normal filament"}.
The background electrons are repetitively rebounded between each filament and undergo backward motion at the filament edges to sustain the return current and stabilize the magnetic vortexes~\cite{SM_animation_normal_pe}. Following the filament coalescence and vortex merging, the field strength grows up to $B_{x,y}\sim 20\,$kT and the electron quantum invariant parameter $\chi_e\equiv(e\hbar/m_e^3c^4)|F_{\mu\nu}p^\nu|$ is close to $\chi_e\approx 0.05$, where $F_{\mu\nu}$ is the field tensor,  $p^\nu$ the electron four-momentum, $m_e$ ($-e$) the electron mass (charge), and $\hbar$ the Planck constant. After experiencing radiative spin flips and transverse deflection by magnetic fields, the beam electrons possess a SP ratio $\langle S_\phi\rangle\approx-3.4\%$ [Fig.~\ref{fig:general}(b)], where $S_\phi\equiv\mathbf{S}\cdot\hat{e}_\phi$ with $\hat{e}_\phi=(-p_y/p_\perp,p_x/p_\perp,0)$ and $p_\perp=(p_x^2+p_y^2)^{1/2}$ represents the SP along the azimuthal direction in the momentum space [Fig.~\ref{fig:general}(e)]. \re{According to Fig.~\ref{fig:general}(f), SP is insignificant for high-energy electrons because of damped radiative spin flips occurring with insufficient synchrotron photon emission.} Therefore, the SP is calculated for electrons within the lowest 5\% energy to filter out the influence of no photon emission.

In contrast, in the ANF case ($n_{pe}=10^{22}\mathrm{cm}^{-3}$), the electron spatial distribution shows a distinct filament structure with background plasmoids encompassed by ultrarelativistic beam electrons [Fig.~\ref{fig:general}(c)].
Topologically, the roles of beam and background electrons are exchanged with each other compared with the NF case, and thus this regime is termed as \textit{``abnormal filament"}.
The key point for the transition from NF to ANF is a counterintuitive feature that the plasmoids gathering of the transversely expelled ions is faster than the pinching of the beam electrons. More simply, the ions respond to the presence of the self-generated plasma fields earlier than the beam electrons.
We may estimate the response time of the ions $t_{pi}^r\sim [m_in_{pe}/|e|^2Z_i n_{be}^2]^{1/2}$~\cite{SM}, and the beam electrons $t_{be}^r\sim[m_e\gamma_{be}/|e|^2n_{be}]^{1/2}$. The criterion $t_{pi}^r\lesssim t_{be}^r$ is equivalent to $n_{pe}\lesssim n_{pe}^\mathrm{a}\equiv  Z_i m_e n_{be}\gamma_{be}/m_i$, where $\gamma_{be}=\varepsilon_{be}/m_ec^2$. 
In the ANF regime, the SP ratio is enhanced to $\langle S_\phi\rangle\approx -10.6\%$ [Figs.~\ref{fig:general}(d)(f)]. While the SP enhancement could be attributed to the increased magnetic field strength $B_{x,y}\approx 34$~kT ($\chi_e\approx 0.06$), 
however, we found that the main reason is different and connected with the deviating dynamics of the filament merging, namely, with the pronounced nonlinear transverse motion of the beam filaments, which we discuss below.

\begin{figure}[t]
\includegraphics[width=1\columnwidth]{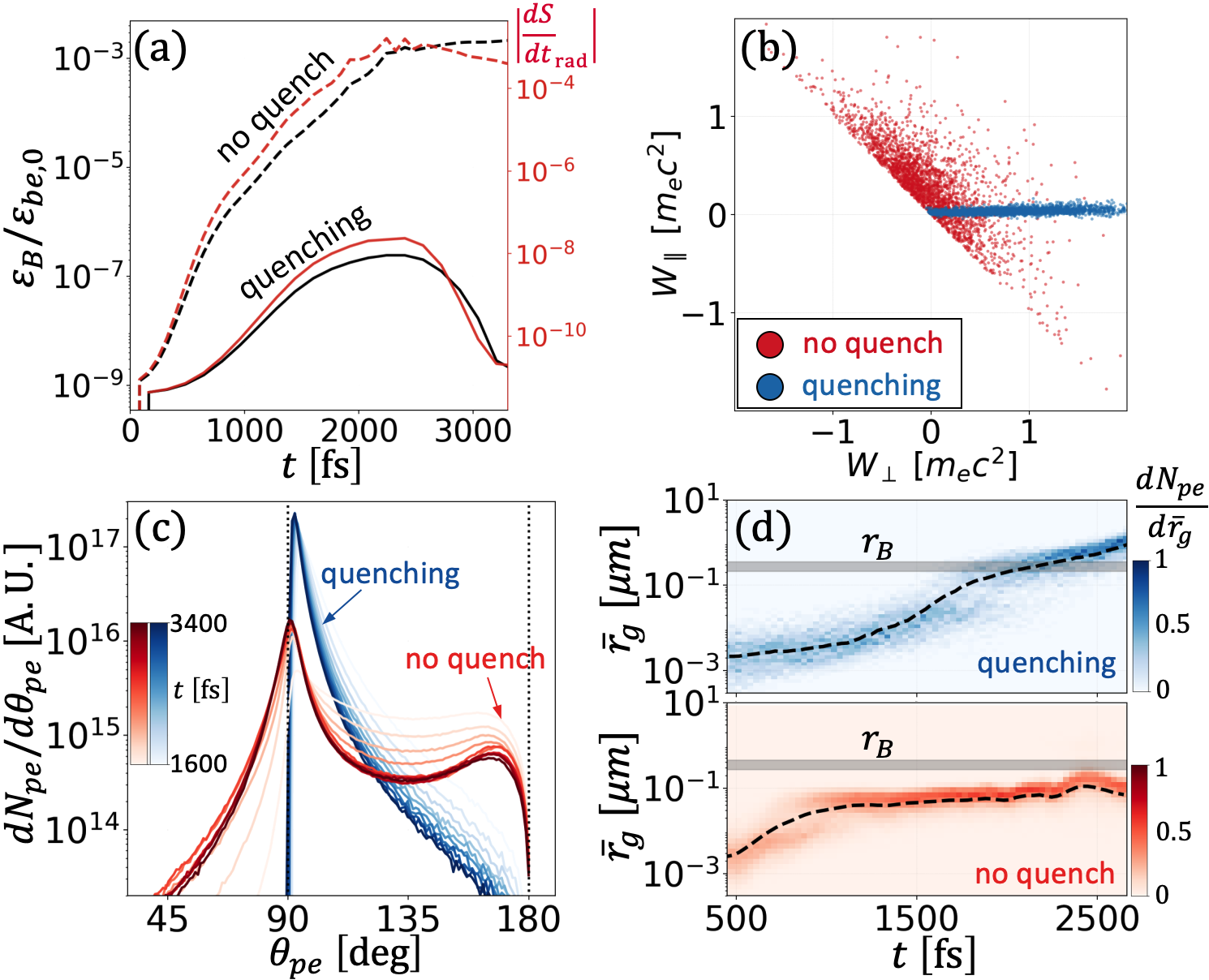}
\caption{(a) Temporal evolution of the magnetic energy ratio $\varepsilon_B/\varepsilon_{be,0}$ and the radiative SP strength $|dS/dt_\mathrm{rad}|$. (b) Work contribution $W_{\parallel,\perp}$ for the background electrons. (c) $dN_{pe}/d\theta_{pe}$ vs $\theta_{pe}$. (d) Time evolution of gyroradius distribution $dN_{pe}/\overline{r}_g$, where the dashed lines show the averaged value and the gray band denotes the scale length $r_B$.}
\label{fig:quenching}
\end{figure}

At high plasma densities the quenching regime sets in [Fig.~\ref{fig:quenching}(a)], where the magnetic field energy $\varepsilon_B$, instead of growing up, declines by two orders of magnitude, which is the reason for the \textit{``quenching"} terming. 
Different from the no quenching situation where the energy gain of background electrons mainly originates from the longitudinal acceleration $W_\parallel=\int -v_zE_z dt$~\cite{SM}, in quenching scenario the background electrons are efficiently accelerated by the transverse sheath field $E_{x,y}$ surrounding the exterior of filaments [Fig.~\ref{fig:quenching}(b)], and thus they are too energetic to be rebounded by the weak magnetic fields. The angular distribution of background electrons $dN_{pe}/d\theta_{pe}$ illustrates that these electrons primarily move along the transverse direction and are no longer deflected backwards to sustain the return current [Fig.~\ref{fig:quenching}(c)]. The unconstrained transverse motion tends to smear out the inhomogeneity of the plasma density and subsequently the magnetic field $B_{x,y}$ is gradually dissipated.
As a result, the radiative SP is drastically reduced [Fig.~\ref{fig:quenching} (a)], with the negligible final SP ratio  $\langle S_\phi\rangle\approx-0.02\%$. We can formulate the quenching criterion by the condition that the gyroradius of background electrons $r_g\sim \gamma_{pe}m_ev_{pe}/|e|B_{x,y}$ is larger than the scale size of the magnetic vortex $r_B\sim 2\pi c/\omega_{pe}$, as validated by the simulation results for the $r_g$ evolution [Fig.~\ref{fig:quenching}(d)]. This condition means, see ~\cite{SM}, that quenching sets in at high plasma densities $n_{pe}\gtrsim n_{pe}^\mathrm{q}\equiv \sqrt{2}\pi^2 \rho n_{be}$, with $\rho\approx 7.2$. 

\begin{figure}
\includegraphics[width=0.9\columnwidth]{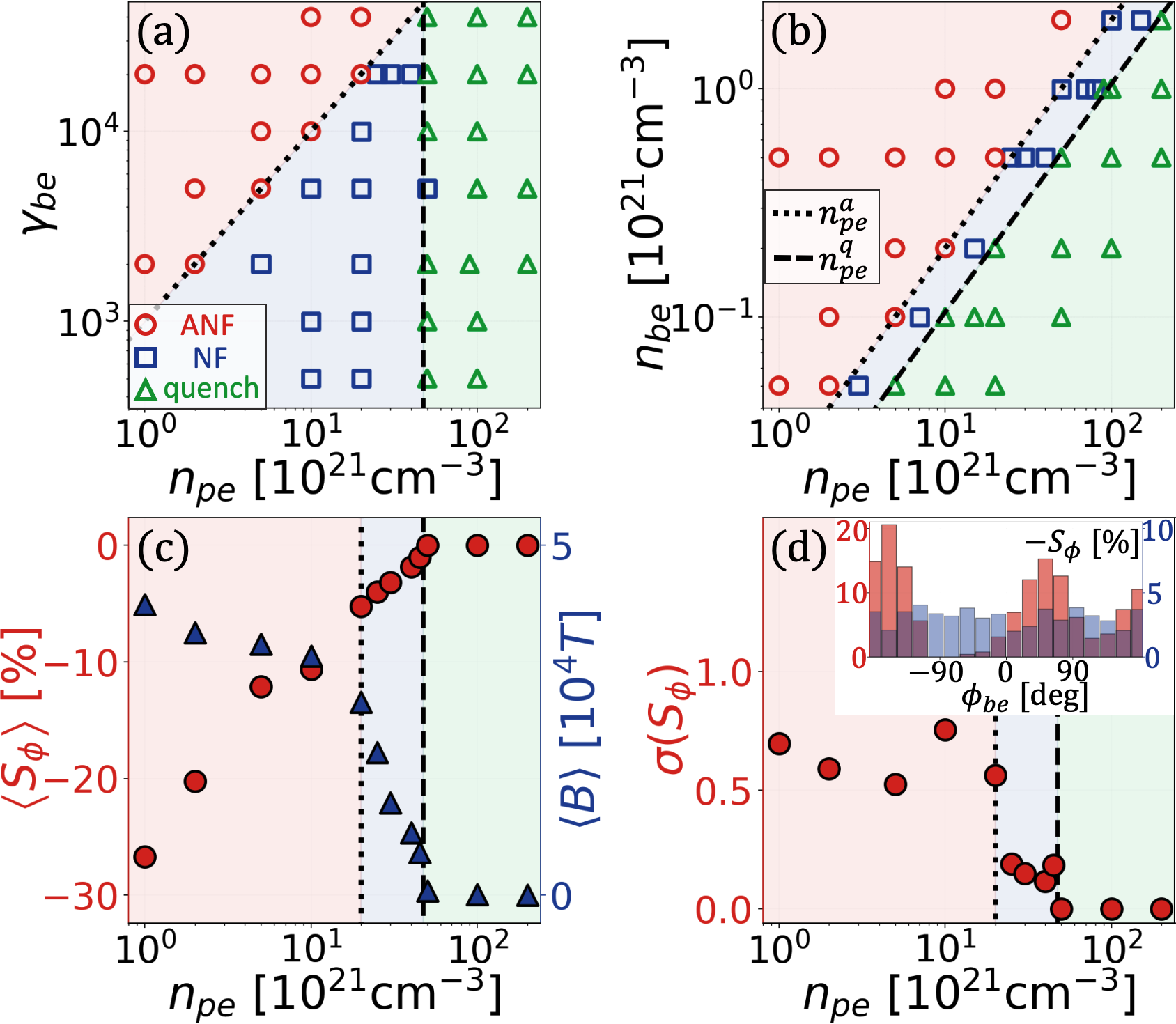}
\caption{\re{Parameter scans in (a) $(n_{pe},\gamma_{be},n_{be}=5\times 10^{20}\mathrm{cm}^{-3})$ and (b) $(n_{pe},n_{be},\gamma_{be}=2\times10^4)$ space, where the markers present the regimes identified by simulations while the regions with red, blue, and green color correspond, respectively, to the ANF, NF, and quenching regimes predicted by $n_{pe}^a$ and $n_{pe}^q$. (c) Dependence of $\langle S_\phi\rangle$ and $\langle B\rangle$ on $n_{pe}$. (d) $\sigma(S_\phi)$ vs $n_{pe}$, where the inset shows the angular distribution of $S_\phi$ for the cases of NF $n_{pe}=3\times10^{21}\mathrm{cm}^{-3}$ (blue) and ANF $10^{21}\mathrm{cm}^{-3}$ (red). In (c)(d), $n_{be}=5\times 10^{20}\mathrm{cm}^{-3}$ and $\gamma_{be}=2\times10^4$. }}
\label{fig:scaling}
\end{figure}

\re{After combining the criterion of ANF and quenching occurrence, the valid ranges of the NF, ANF, and quenching regimes are derived as $n_{pe}^\mathrm{a}< n_{pe} < n_{pe}^\mathrm{q}$, $n_{pe}\lesssim n_{pe}^\mathrm{a}$, and $n_{pe} \gtrsim n_{pe}^\mathrm{q}$, respectively. These criteria are proven by the parameter scans of simulations in $(n_{pe},\gamma_{pe})$ and $(n_{pe},n_{be})$ space [Figs.~\ref{fig:scaling}(a)(b)].
To exhibit the distinct SP properties in the three regimes, we change the background plasma density $n_{pe}$ while fixing the beam electron parameters $n_{be}$ and $\gamma_{be}$. The variation of the SP ratio $\langle S_\phi\rangle$ and effective magnetic field $\langle B\rangle$ versus $n_{pe}$ in Fig.~\ref{fig:scaling}(c) demonstrates:} i) $\langle S_\phi\rangle$ and $\langle B\rangle$ are negligible in the quenching regime; ii) Both $\langle S_\phi\rangle$ and $\langle B\rangle$ exhibit linear variation tendency in the NF regime; iii) In ANF, $\langle S_\phi\rangle$ increases but $\langle B\rangle$ stays nearly unchanged when $n_{pe}$ decreases. The latter property indicates that the stronger magnetic field in ANF with respect to NF cannot solely explain the larger SP.
Another distinct feature between ANF and NF regimes is the inhomogeneity of the SP angular distribution $S_\phi$ vs $\phi_{be}$ [$\phi_{be}=\arctan2(p_y,p_x)$]. The inhomogeneity quantified by the dispersion of the angular distribution  $\sigma(S_\phi)$~\cite{SM_inhomogeneity} is significantly larger in the ANF regime [Figs.~\ref{fig:scaling}(d)].

\begin{figure}
\includegraphics[width=1\columnwidth]{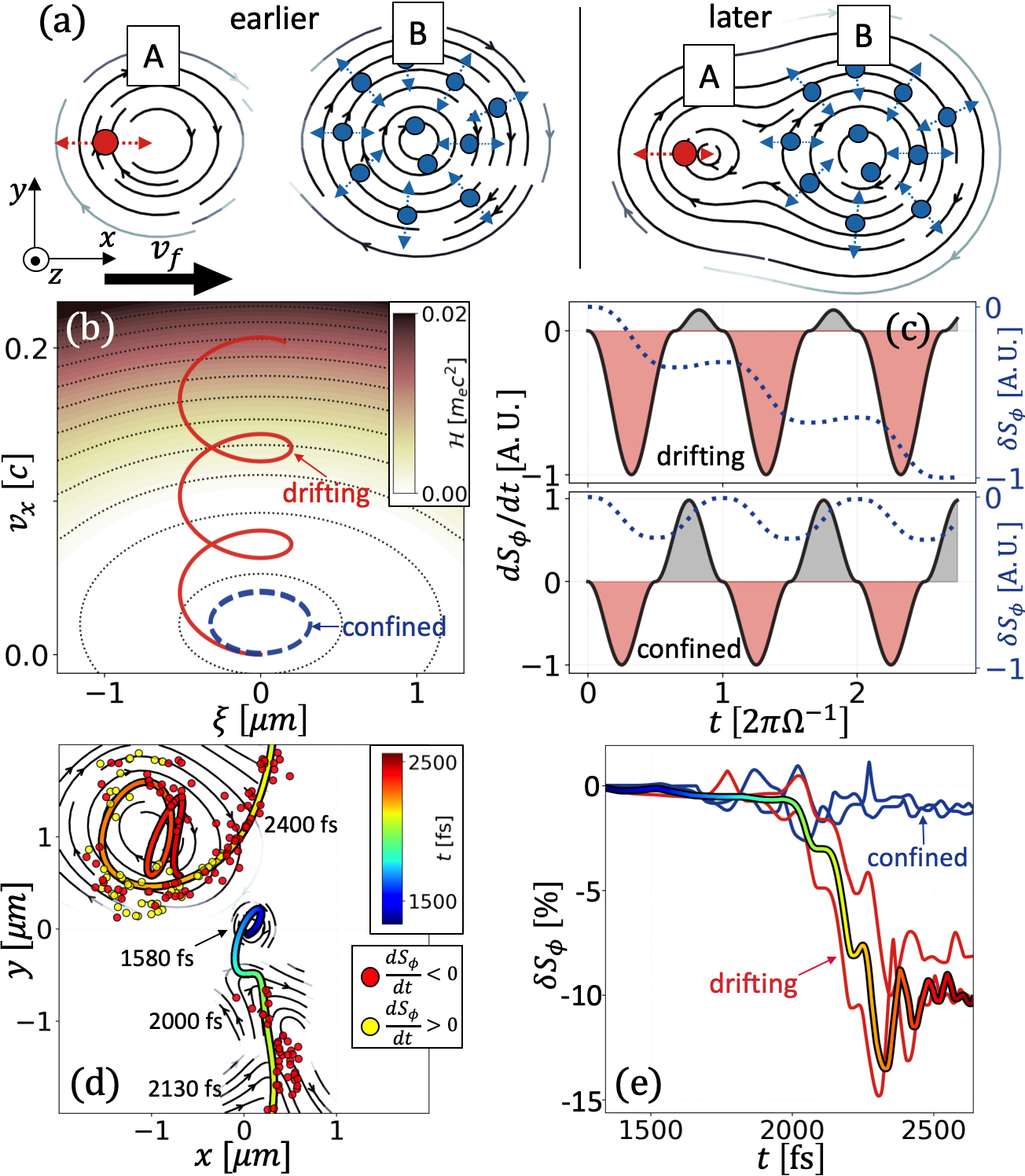}
\caption{(a) Schematic of drifting (red) and confined (blue) electron dynamics, where the vortexes present the magnetic field. (b) Electron trajectories with $a_0=10^{-4}>a_0^\mathrm{s}$ (red) and $10^{-6}<a_0^\mathrm{s}$ (blue) in $(\xi,v_x)$. (c) Time evolution of $dS_\phi/dt_\mathrm{rad}$ (black) and $\delta S_\phi$ (blue). In (b)(c) the parameters are $\dot{\xi}_0 =0.002$, $\gamma_{be}=10^4$, and $\kappa_B=0.5$, i.e. $a_0^\mathrm{s}\approx 1.4\times 10^{-5}$. (d) A drifting electron trajectory (in rainbow colorcode), where streamlines present the magnetic field at different time while the red/yellow dots refer to photon emission accompanied with negative/positive spin flips. (e) Temporal evolution of $\delta S_\phi$ for drifting (red) and confined (blue) electrons.}
\label{fig:hami_dym}
\end{figure}

Returning to the question of the high SP in ANF, during filament merging in the ANF, the topological connectedness of the ultrarelativistic electron flow leads to a pronounced nonlinear transverse motion of the beam filaments because of the lack of the impediment of the background plasma in contrast to NF.
This transverse motion is critical to enhance the SP ratio. We have developed an 1D Hamiltonian model to analyze this effect [Fig.~\ref{fig:hami_dym}(a)]. Assume the transverse velocity of the filament is $v_f$, and the magnetic vortex field exerted on the electron $B_y(x-v_ft)$. In the vortex's co-moving frame $\xi\equiv x-v_ft$, the electron dynamics is characterized by the equations $\dot{\xi} = v_x-v_f$  and $\ddot{\xi} = -d\Psi(\xi)/d\xi -a_0$, where $a_0\equiv \dot{v}_f$ is the vortex's acceleration, and $\dot{\xi}$ the relative velocity of the electron inside the magnetic vortex (the overdot is a time derivative). $\Psi(\xi)=-\int |e|cB_y(\xi)/\gamma_{be} d\xi$ is the potential of the magnetic vortex field, given $v_z\approx c$. 
The equation of motion above can be reformulated as the conserved Hamiltonian:
\begin{eqnarray}\label{eq:hami_H}
\mathcal{H}(\xi,\dot{\xi})=\frac{1}{2}\dot{\xi}^2+\Psi(\xi)+a_0\xi.
\end{eqnarray}
Eq.~(\ref{eq:hami_H}) describes the electron oscillatory dynamics: $d\xi/dt = \pm \sqrt{2[2a_0(\xi-\xi_0)-\Psi(\xi)-\dot{\xi_0}^2]}$~\cite{SM}, where $\xi=\xi_0$ and $\dot{\xi}=\dot{\xi}_0$ are the initial conditions at $t=0$.
Assuming $B_y(\xi)=-\kappa_B\xi$ (in accordance with the simulation results), and $\Psi(\xi)=\Omega^2\xi^2/2$ with $\Omega\equiv\sqrt{\kappa_B|e|c/\gamma_{be}}$, the oscillatory motion within the  magnetic vortex is
\begin{eqnarray}\label{eq:xi_xiv}
\xi&=&\dot{\xi_0}\frac{1}{\Omega}\sin\Omega t + (\xi_0+\frac{a_0}{\Omega^2})\cos\Omega t -\frac{a_0}{\Omega^2},
\end{eqnarray}
featuring either confined ($a_0\rightarrow 0$) or drifting motion [Fig.~\ref{fig:hami_dym}(a)(b)]. This oscillatory dynamics is accompanied by photon emissions and by consequent radiative spin flips and radiative polarization of electrons.

Based on Eq.\eqref{eq:xi_xiv}, the evolution of electron SP is calculated as~\cite{SM}
\begin{eqnarray}\label{eq:dS_phi_dt}
\frac{dS_\phi}{dt}_\mathrm{rad}\approx-\frac{\sqrt{3}\alpha_f m_ec^2}{h}\frac{\chi_e}{\gamma_e}\mathcal{W}(\chi_e)\frac{B_y(\xi)}{|B_y(\xi)|}\frac{v_x}{|v_x|},
\end{eqnarray}
where $\mathcal{W}(\chi_e)=\int  2\chi_{ph}^2/[3\chi_e^3(\chi_e-\chi_{ph})]K_{1/3}(u) d\chi_{ph}$, $u=2\chi_{ph}/[3\chi_e(\chi_e-\chi_{ph})]$, and $K_{1/3}$ is the modified secondary Bessel function. 
The term $\Pi\equiv (B_y/|B_y|)(v_x/|v_x|)$ characterizing the SP reads approximately (at $\xi_0=0$ and $v_f=v_{f0}$ at $t=0$):
\begin{eqnarray}\label{eq:symmetry}
\Pi \propto
& \mathcal{O}(t) + a_0(\frac{\dot{\xi}_0}{\Omega}t\sin\Omega t +\frac{a_0}{\Omega^2}t\cos\Omega t - \frac{a_0t}{\Omega^2}),
\end{eqnarray}
where $\mathcal{O}(t)$ is periodic with frequency $\Omega$ and does not lead to a net average SP (usually the case for the NF regime).
The second term in Eq.\eqref{eq:symmetry} linearly proportional to $t$ yields net SP when the filament transverse acceleration is large $a_0\gtrsim a_0^s \equiv \dot{\xi_0}\Omega$, which is the case in the ANF regime. While for the confined case with $a_0<a_0^\mathrm{s}$, the SP gain and loss compensate each other due to the nearly symmetric spin flips~\cite{SM_animation_Hami_confine}, for the drifting case with  the moving magnetic vortex $a_0>a_0^\mathrm{s}$  the oscillations and spin flips are not symmetric~\cite{SM_animation_Hami_drift}, which leads to the pronounced net SP gain [Fig.~\ref{fig:hami_dym}(c)].

\begin{figure}[t]
\includegraphics[width=0.78\columnwidth]{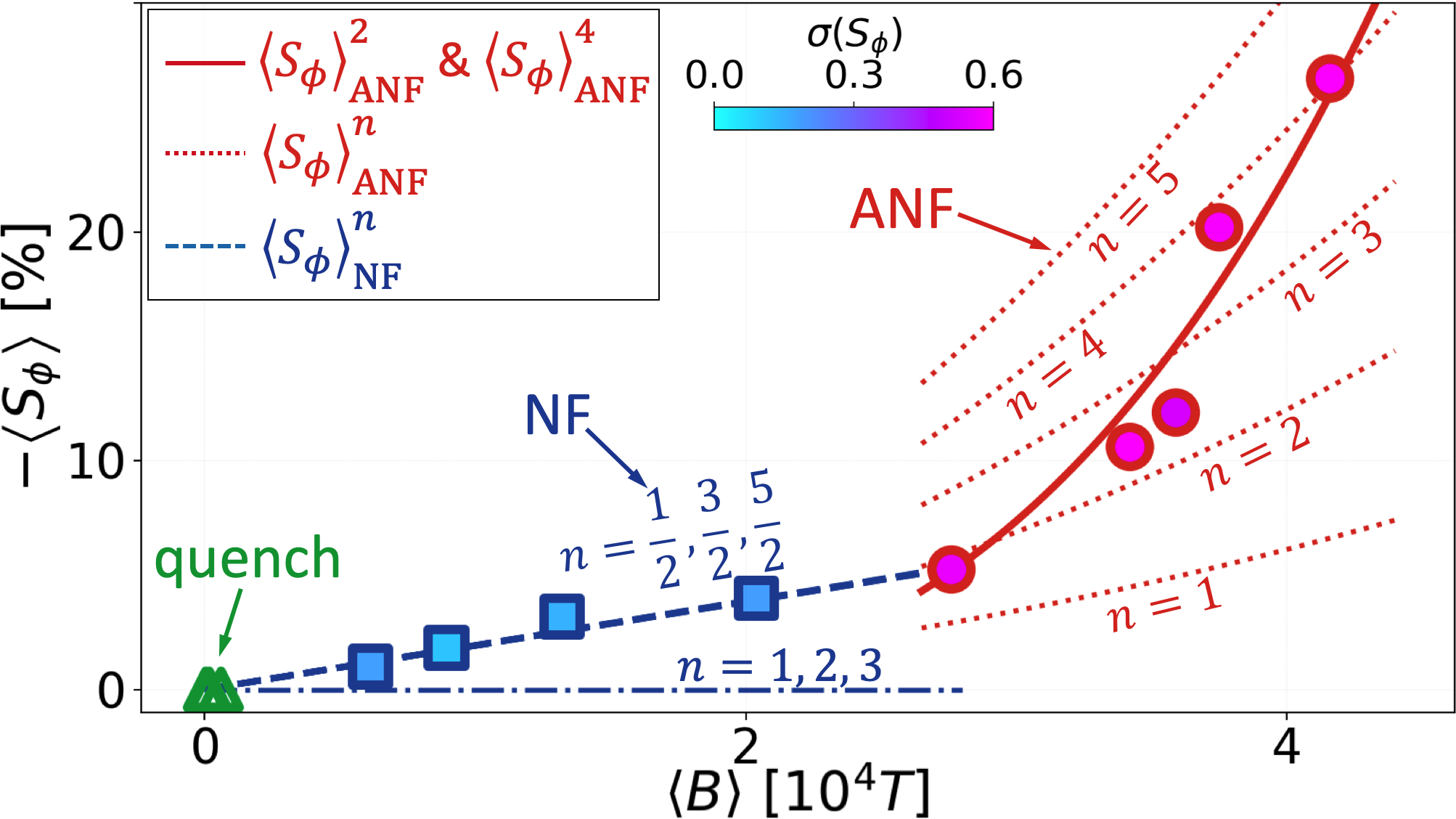}
\caption{Correlation between $\langle S_\phi\rangle$ and $\langle B\rangle$, where the color filled in the markers denotes $\sigma(S_\phi)$. 
}
\label{fig:svariance}
\end{figure}

The Hamiltonian analysis is confirmed by the simulation results of the ANF case. Figure~\ref{fig:hami_dym}(d) presents a representative electron comoving with the magnetic vortex, where the photon emission accompanied by a negative spin flip $dS_\phi/dt<0$ dominates the whole emission procedure~\cite{SM_animation_abnormal_be}. The electron SP gradually rises up to $\delta S_\phi \approx -10\%$ [Fig.~\ref{fig:hami_dym}(e)]. Note that there are confined electrons existing in the ANF case as well since some plasma filaments move too slowly to satisfy $a_0\gtrsim a_0^\mathrm{s}$.

The correlation between $\langle S_\phi\rangle$ and $\langle B\rangle$ is shown in Fig.~\ref{fig:svariance}. The slope of the correlation in the ANF is distinct from NF, which stems from the net SP $\delta S_\phi$ accumulated by the different periods of the electron oscillation in the drifting magnetic vortex in ANF. This correlation can be explained by a combination of $\langle S_\phi\rangle_\mathrm{ANF}^2$ and $\langle S_\phi\rangle_\mathrm{ANF}^4$ [Fig.~\ref{fig:svariance}], where $\langle S_\phi\rangle_\mathrm{ANF}^n =\int_0^{n(2\pi/\Omega)}dS_\phi/dt_\mathrm{rad} dt$ is the numerical integral of Eq.~\eqref{eq:dS_phi_dt} and the integer $n$ denotes the number of oscillation periods. 
In contrast, the SP gain $\delta S_\phi$ in the NF originates from electrons' uncompensated half period oscillations while the one with integer period would lead to $\delta S_\phi\approx 0$. Considering $\chi_e\approx 0.05$ and $\mathcal{W}(\chi_e)\approx 0.5\chi_e^{3/2} \approx0.006$ obtained from the simulation results in the NF cases, the radiative SP ratio can be estimated as 
$\langle S_\phi\rangle_\mathrm{NF}^n\sim \sqrt{3}\eta\alpha_f/(2\pi)\sqrt{\gamma_{be}/m_en_{be}}\mathcal{W}(\chi_e)\left<B\right>$ with the coefficient $\eta\approx 0.4$ accounting for the nonuniform amplitude in integration~\cite{SM}. 
The discrepancy between $\langle S_\phi\rangle_\mathrm{NF}^n$ and $\langle S_\phi\rangle_\mathrm{ANF}^n$ underlines the significant influence of the nonlinear transverse drifting motion of filaments on the electron SP.

\re{Additional 3D simulations are performed and the results qualitatively reproduce those of the 2D simulations, indicating that the plasma motion along the longitudinal direction merely plays a secondary role in influencing the features of the ultrarelativistic CFI and electron SP.} 
The kinetic mechanisms of CFI regimes identified here have valuable implications for both laboratory and astrophysical phenomena. As a concrete application, the filaments' nonlinear transverse motion identified in the ANF regime can be harnessed to compress in time the photon emission and in this way to improve the peak brilliance of gamma-ray sources in~\cite{benedetti2018giant}.
In astrophysics, the fast ion motion discerned in the ANF regime may influence the internal structure of collisionless shocks~\cite{lobet2015ultrafast,bykov2011fundamentals}.
The ultrafast coalescence dynamics characteristic for ANF could enable the magnetic reconnection to drive stellar flares~\cite{yamada2010magnetic,ji2022magnetic} and work as scattering magnetic bodies to modulate the cosmic ray's transportation~\cite{fermi1949origin,bell2013cosmic,caprioli2013cosmic}. 
Furthermore, the ANF regime also manifests that an asymmetric field is no longer a necessity for producing spin-polarized plasmas, implying the intrinsic existence of electron SP in the fast cooling stage during a gamma-ray burst~\cite{gamma_ray_burst}.

In conclusion, we have studied radiative SP of ultrarelativistic beam electrons in plasma CFI. Through particle-in-cell simulations, three different current filament regimes (NF, ANF, and quenching) could be qualitatively distinguished via the combined information of $\langle S_\phi\rangle$ and angular inhomogeneity $\sigma(S_\phi)$, which implies the potential of electron SP to serve as a new information source in investigating laboratory and astrophysical plasma instabilities. 

The original version of code EPOCH adapted here is funded by the UK EPSRC grants EP/G054950/1, EP/G056803/1, EP/G055165/1 and EP/ M022463/1. The authors would like to thank Laurent Gremillet, Matteo Tamburini, and Sergei Bulanov for the fruitful discussion regarding plasma stream instabilities. Z. G. also thanks Zhi-Qiu Huang for the discussion about astrophysical collisionless shocks and gamma-ray bursts.


\input{output.bbl}
\end{document}

%% file: output.bbl
%

%% file: main.bbl
\begin{thebibliography}{86}
\bibitem [{SM_()}]{SM_inhomogeneity}%
  \BibitemOpen
  \href@noop {} {}\bibinfo {howpublished} {This inhomogeneity can be quantified
  as the standard deviation
  $\sigma(S_\phi)=\sqrt{\mathrm{E}[(S_{\phi,i}-\mathrm{E}[S_{\phi,i}])^2]}$,
  where $S_{\phi,i}$ is the value at each angle bin and $\mathrm{E}[...]$ the
  expectation. The dependence of $\sigma(S_\phi)$ on plasma density $n_{pe}$
  indicates that the SP angular distribution tends to be more
  inhomogeneous/uniform in the ANF/NF regimes.}\BibitemShut {Stop}%
\end{thebibliography}
